\documentstyle[preprint,prl,aps,epsfig]{revtex}
\begin{document}
\title{First- principle calculations of magnetic interactions in correlated systems}
\author{M. I. Katsnelson$^1$ and A. I. Lichtenstein$^2$}
\address{
$^1$ Institute of Metal Physics, 620219 Ekaterinburg , Russia \\
$^2$ Max-Planck-Institut f\"ur Festk\"orperforschung, 70569 Stuttgart, Germany }

\maketitle

\begin{abstract}
We present a novel approach to calculate the effective exchange interaction
parameters based on the realistic electronic structure of correlated
magnetic crystals in local approach with the frequency dependent self
energy. The analog of ``local force theorem'' in the density functional
theory is proven for highly correlated systems. The expressions for
effective exchange parameters, Dzialoshinskii- Moriya interaction, and
magnetic anisotropy are derived. The first-principle calculations of
magnetic excitation spectrum for ferromagnetic iron, with the local
correlation effects from the numerically exact QMC-scheme is presented.
\end{abstract}

\section{Introduction}

The calculation of thermodynamic properties and excitation spectra of
different magnetic materials is one of the most important motivations of the
microscopic theory of magnetism. The main approach for such type of
investigations is the local spin density functional (LSDF) scheme \cite{lsdf}%
. However, this method has some serious shortcomings when applying to
transition metal and rare-earth magnetic materials. The main defect is the
absence of the ``Hubbard'' type correlations which are most important for
real magnets (see, e.g., recent reviews \cite{pwa,vons,dmft}). This leads
to, generally speaking, incorrect description of electronic structure for
such important groups of magnetic materials as rare-earth metals and their
compounds, metal-oxide compounds (including ``classical'' Mott insulators
such as NiO and MnO as well as high-$T_c$ cuprates) and even for the
iron-group metals \cite{pwa,dmft,lda++}. At the same time, the experience
with the Hubbard model shows that the description of electronic structure
and magnetic properties of highly correlated materials are closely
connected. Recently we propose a rather general scheme (so-called ``LDA++
approach'') for first-principle calculations of the electronic structure
with the local correlation effects being included \cite{lda++}. In this
technique the full matrix of on-site Coulomb repulsion for the correlated
states is taken into account in the local approximation for the electron
self- energy. In such a way, we could provide a rather reasonable
description of the electronic structure for different correlated systems
such as Fe, NiO, and TmSe. It will be very useful to develop this
approach for the description of different magnetic characteristics.

The most rigorous way to consider properties of magnetic excitations is the
calculation of frequency-dependent magnetic susceptibilities\cite{dmft}.
However, for many important cases we can restrict ourselves to more simple
problem of the calculation of {\it static }response functions. More
explicitly, one can consider the variations of total energy $E_{tot}$ (or
thermodynamic potential $\Omega $) with respect to magnetic moments
rotations. This approach results in the magnetic interactions of different
types: the variation of total energy of a ferromagnet over the rotation of
all spins at the same angle determines the magnetic anisotropy energy, while
the variation of $E_{tot}$ over the relative rotations of spins on two sites
gives the parameters of pairwise exchange interactions, etc. This approach
was proposed earlier in the the framework of the LSDF scheme \cite{LKG}. It
is sufficient for the calculation of ``phenomenological'' exchange
parameters which are important for the consideration of domain wall widths
and other ``micromagnetic'' properties. In the adiabatic approximation when
the energies of magnetic excitations are small in comparison with typical
electronic energies this is also sufficient for the calculation of the
spin-wave spectrum. In the mean-field approximation these ``exchange
parameters'' can be used for the estimation of Curie or Neel temperature.

In this work we derive general expressions for the parameters of magnetic
interactions in LDA++ approach and calculate the exchange parameters for
ferromagnetic iron. It is a first attempt to investigate magnetic
interactions, taking into account correlation effects in the electronic
structure for real materials.

\section{General formalism}

\subsection{Local force theorem in LDA++ approach}

An important trick for the definition of exchange interactions in the LSDF
approach is the use of so called ''local force theorem''. This reduces the
calculation of the total energy change to the variations of one-particle
density of states \cite{MA,LKG}. First of all, let us prove the analog of
local force theorem in the LDA++ approach. In contrast with the standard
density functional theory, it deals with the real dynamical quasiparticles
defined via Green functions for the correlated electrons rather than with
Kohn-Sham ``quasiparticles'' which are, strictly speaking, only auxiliary
states to calculate the total energy. Therefore, instead of the working with
the thermodynamic potential $\Omega $ as a {\it density} functional we have
to start from its general expression in terms of an exact Green function 
\cite{luttinger}

\begin{eqnarray}
\Omega &=&\Omega _{sp}-\Omega _{dc}  \label{first} \\
\Omega _{sp} &=&-Tr\left\{ \ln \left[ \Sigma -G_0^{-1}\right] \right\} 
\nonumber \\
\Omega _{dc} &=&Tr\Sigma G-\Phi  \nonumber
\end{eqnarray}
where $G,G_0$ and $\Sigma $ are an exact Green function, its bare value and
self-energy, correspondingly; $\Phi $ is the Luttinger generating functional
(sum of the all connected skeleton diagrams without free legs), $%
Tr=Tr_{\omega iL\sigma }$ is the sum over Matsubara frequencies $Tr_\omega
...=T\sum\limits_\omega ...,$ $\omega =\pi T\left( 2n+1\right) ,$ $n=0,\pm
1,...,$ $T$ is the temperature, and $iL\sigma $ are site numbers ($i$),
orbital quantum numbers ($L={l,m}$) and spin projections $\sigma $ ,
correspondingly. We have to keep in mind also Dyson equation 
\begin{equation}
G^{-1}=G_0^{-1}-\Sigma  \label{DYSON}
\end{equation}
and the variational identity

\begin{equation}
\delta \Phi =Tr\Sigma \delta G.  \label{var}
\end{equation}
We represent the expression (\ref{first}) as a difference of ''single
particle'' ($sp$) and ''double counted'' ($dc$) terms as it is usual in the
density functional theory. When neglecting the quasiparticle damping, $%
\Omega _{sp}$ will be nothing but the thermodynamic potential of ''free''
fermions but with exact quasiparticle energies. Suppose we change the
external potential, for example, by small spin rotations. Then the variation
of the thermodynamic potential can be written as 
\begin{equation}
\delta \Omega =\delta ^{*}\Omega _{sp}+\delta _1\Omega _{sp}-\delta \Omega
_{dc}  \label{var2}
\end{equation}
where $\delta ^{*}$ is the variation without taking into account the change
of the ''self-consistent potential'' (i.e. self energy) and $\delta _1$ is
the variation due to this change of $\Sigma $. Taking into account Eq. (\ref
{var}) it can be easily shown (cf. \cite{luttinger}) that 
\begin{equation}
\delta _1\Omega _{sp}=\delta \Omega _{dc}=TrG\delta \Sigma  \label{var3}
\end{equation}
and hence 
\begin{equation}
\delta \Omega =\delta ^{*}\Omega _{sp}=-\delta ^{*}Tr\ln \left[ \Sigma
-G_0^{-1}\right]  \label{var4}
\end{equation}
which is an analog of the ''local force theorem'' in the density functional
theory \cite{LKG}. In the LSDF scheme all the computational results
expressed in terms of the retarded Green function and not in the Matsubara
one. The relations of ``real'' and ``complex'' Green-function formulae are
based on the identity 
\begin{equation}
Tr_\omega F(i\omega )=-\frac 1\pi \int\limits_{-\infty }^\infty
dzf(z)ImF(z+i0)  \label{id}
\end{equation}
where $f(z)=\left[ \exp z/T+1\right] ^{-1}$ is the Fermi function, $F(z)$ is
a function regular in all the complex plane except real axis. Therefore Eq.(%
\ref{var4}) takes the following form 
\begin{equation}
\delta \Omega =\frac 1\pi \int\limits_{-\infty }^\infty dzf(z)ImTr_{iL\sigma
}\ln G^{-1}(z+i0)  \label{lloyd}
\end{equation}
which is the starting point for the calculations of magnetic interactions in
LSDF approach \cite{LKG}. However note, that in the case of
frequency-dependent self-energy (LDA++ approach) it is more convenient to
work with the Matsubara Green functions \cite{lda++}.

\subsection{Effective spin Hamiltonian}

Further considerations are similar to the corresponding ones in LSDF
approach. The most suitable way based on the sum rule is proposed in \cite
{physica}.

In the LDA++ scheme, the self energy is local, i.e. is diagonal in site
indices. Let us write the spin-matrix structure of the self energy and Green
function in the following form 
\begin{eqnarray}
\Sigma _i &=&\Sigma _i^c+{\bf \Sigma}_i^s{\bf { \sigma }}  \label{spin} \\
G_{ij} &=&G_{ij}^c+{\bf G}_{ij}^s{\bf { \sigma }}  \nonumber
\end{eqnarray}
where $\Sigma _i^{\left( c,s\right) }=\frac 12\left( \Sigma _i^{\uparrow
}\pm \Sigma _i^{\downarrow }\right)$, ${\bf \Sigma}_i^s=\Sigma _i^s{\bf e}_i,
$ with ${\bf e}_i$ being the unit vector in the direction of effective
spin-dependent potential on site $i$, ${\bf { \sigma }}=(\sigma_x,\sigma_y,%
\sigma_z)$ are Pauli matrices, $G_{ij}^c=\frac 12Tr_\sigma (G_{ij})$ and $%
{\bf G}_{ij}^s=\frac 12Tr_\sigma (G_{ij} {\bf {\sigma}})$. We assume that
the bare Green function $G^0$ does not depend on spin directions and all the
spin-dependent terms including the Hartree-Fock terms are incorporated in
the self energy. Spin excitations with low energies are connected with the
rotations of vectors ${\bf e}_i$:

\begin{equation}
\delta {\bf e}_i=\delta {\bf \varphi}_i\times {\bf e}_i  \label{rot1}
\end{equation}
According to the ''local force theorem'' (\ref{var4}) the corresponding
variation of the thermodynamic potential can be written as 
\begin{equation}
\delta \Omega =\delta ^{*}\Omega _{sp}={\bf V}_i\delta {\bf \varphi}_i
\label{rot2}
\end{equation}
where the torque is equal to 
\begin{equation}
{\bf V}_i=2Tr_{\omega L}\left[ {\bf \Sigma }_i^s\times {\bf G}_{ii}^s\right]
\label{torque}
\end{equation}
Further we have to use an important sum rule for the Green function which is
the consequence of Dyson equation. Using Eq.(\ref{DYSON}) one has 
\begin{eqnarray}
G &=&G^c\cdot \widehat{I}+{\bf G}^s\cdot {\bf {\sigma}}=GG^{-1}G=
\label{sum1} \\
&&\ \ G\left( \left( G_0^{-1}-\Sigma ^c\right) \cdot \widehat{I}-{\bf \Sigma}%
^s\cdot {\bf {\sigma}}\right) \left( G^c\cdot \widehat{I}+{\bf G}^s\cdot 
{\bf {\sigma}}\right) .  \nonumber
\end{eqnarray}
Separating the spin-dependent and the spin-independent parts in Eq. (\ref
{sum1}) we have the following sum rules for $G^c$%
\begin{equation}
G^c=G^c\left( G_0^{-1}-\Sigma ^c\right) G^c-G^c{\bf \Sigma}^s{\bf G}%
^s=G^c\left( G_0^{-1}-\Sigma ^c\right) G^c-{\bf G}^s{\bf \Sigma}^sG^c
\label{sum2}
\end{equation}
and similarly for ${\bf G}^s$%
\begin{eqnarray}
{\bf G}^s &=&-\left( {\bf G}^s{\bf \Sigma}^s\right) {\bf G}^s+G^c{\bf \Sigma}%
^sG^c+iG^c\left( {\bf \Sigma}^s\times{\bf G}^s\right) =  \label{sum3} \\
\ &=&-{\bf G}^s\left( {\bf \Sigma}^s{\bf G}^s\right)+G^c{\bf \Sigma}%
^sG^c+i\left( {\bf G}^s\times{{\bf \Sigma} }^s\right) G^c  \nonumber
\end{eqnarray}
Then for diagonal elements of the Green function one obtains 
\begin{equation}
{\bf G}_{ii}^s=-\sum_j\left[ \left( {\bf G}_{ij}^s{\bf \Sigma }_j^s\right) 
{\bf G}_{ji}^s-G_{ij}^c{\bf \Sigma }_j^sG_{ji}^c-iG_{ij}^c\left( {\bf \Sigma 
}_j^s\times {\bf G}_{ji}^s\right) \right]  \label{sum4}
\end{equation}
Substituting (\ref{sum4}) into (\ref{torque}) we have the following
expression for the torque 
\begin{eqnarray}
{\bf V}_i &=&2Tr_{\omega L}\left[ {\bf \Sigma }_i^s\times {\bf G}_{ii}\right]
\label{torque2} \\
\ &=&-2\sum_jTr_{\omega L}\left\{ {\bf \Sigma }_i^s\times \left( {\bf G}%
_{ij}^s{\bf \Sigma }_j^s\right) {\bf G}_{ji}^s-{\bf \Sigma }_i^s\times
G_{ji}^c{\bf \Sigma }_j^sG_{ij}^c-i{\bf \Sigma}_i^s\times G_{ji}^c\left( 
{\bf \Sigma }_j^s\times {\bf G}_{ji}^s\right) \right\}  \nonumber
\end{eqnarray}
If we represent the total thermodynamic potential of spin rotations or the
effective Hamiltonian in the form 
\begin{equation}
\Omega _{spin}=-\sum_{ij}Tr_{\omega L}\left\{ \left( {\bf G}_{ij}^s{\bf %
\Sigma }_j^s\right) \left( {\bf G}_{ji}^s{\bf \Sigma }_i^s\right) -{\bf %
\Sigma }_i^sG_{ij}^c{\bf \Sigma }_j^sG_{ji}^c-i\left( {\bf \Sigma }%
_i^s\times G_{ij}^c{\bf \Sigma }_j^s\right) {\bf G}_{ji}^s\right\}
\label{Hamilt}
\end{equation}
one can show by direct calculations that 
\begin{equation}
\left[ \frac{\delta \Omega _{spin}}{\delta {\bf \varphi }_i}\right]
_{G=const}={\bf V}_i  \label{torque3}
\end{equation}
This means that $\Omega _{spin}\left\{ {\bf e}_i\right\} $ is the effective
spin Hamiltonian. The last term in Eq.(\ref{Hamilt}) is nothing but
Dzialoshinskii- Moriya interaction term. It is non-zero only in relativistic
case where ${\bf \Sigma }_j^s$ and ${\bf G}_{ji}^s$ can be, generally
speaking, ``non-parallel'' and $G_{ij}\neq G_{ji}$ for the crystals without
inversion center. In the following we will not consider this term.

\subsection{Exchange interactions}

In the nonrelativistic case one can rewrite the spin Hamiltonian for small
spin deviations near collinear magnetic structures in the following form 
\begin{equation}
\Omega _{spin}=-\sum_{ij}J_{ij}{\bf e}_i{\bf e}_j  \label{heis}
\end{equation}
where 
\begin{equation}
J_{ij}=-Tr_{\omega L}\left( \Sigma _i^sG_{ij}^{\uparrow }\Sigma
_j^sG_{ji}^{\downarrow }\right)  \label{Jij}
\end{equation}
are the effective exchange parameters. This formula generalize the LSDA
expressions of \cite{LKG} to the case of correlated systems.

The sum rule (Eq. \ref{sum4}) for the collinear magnetic configuration takes
the following form 
\begin{equation}
G_{ii}^{\uparrow }-G_{ii}^{\downarrow }=2\sum\limits_jG_{ij}^{\uparrow
}\Sigma _j^sG_{ji}^{\downarrow }  \label{sum5}
\end{equation}
Using Eq. (\ref{sum5}) we obtain the following expression for the total
exchange interaction of a given site with the all magnetic environment. 
\begin{equation}
J_i=\sum\limits_{j(\neq i)}J_{ij}=Tr_{\omega L}\left[ \Sigma
_i^sG_{ii}^{\uparrow }\Sigma _i^sG_{ii}^{\downarrow }-\frac 12\Sigma
_i^s\left( G_{ii}^{\uparrow }-G_{ii}^{\downarrow }\right) \right]  \label{Jo}
\end{equation}

Spin wave spectrum in ferromagnets can be considered both directly from the
exchange parameters or by the consideration of the energy of corresponding
spiral structure (cf \cite{LKG}). In nonrelativistic case when the
anisotropy is absent one has 
\begin{equation}
\omega _{{\bf q}}=\frac 4M\sum\limits_jJ_{0j}\left( 1-\cos {\bf qR}_j\right)
\equiv \frac 4M[J(0)-J({\bf q})]  \label{om}
\end{equation}
where $M$ is the magnetic moment (in Bohr magnetons) per magnetic ion.
Corresponding expressions can be easily written in {\bf k-}space also. In
the short notation $q=({\bf q},i\omega )$ it is easy to write the general
expression for $J({{\bf q)}\equiv }J({\bf q},0)$:

\begin{equation}
J(q)=-\frac 1{N_k}\sum_kTr_L[\Sigma ^s(k)G^{\uparrow }(k)\Sigma
^s(k+q)G^{\downarrow }(k+q)]  \label{Jq}
\end{equation}
where $N_k$ is the total number of k-points.

It should be noted that the expression for spin stiffness tensor $D_{\alpha
\beta }$ defined by the relation  
\begin{equation}
\omega _{{\bf q}}=D_{\alpha \beta }q_\alpha q_\beta   \label{D}
\end{equation}
(${\bf q\rightarrow 0}$) in terms of exchange parameters has to be exact
since it is the consequence of phenomenological Landau- Lifshitz equations
which are definitely correct in the long-wavelength limit (cf \cite{LKG}).
One has from Eqs. (\ref{Jij},\ref{om}) 
\begin{equation}
D_{\alpha \beta }=\frac 2M\sum\limits_jJ_{0j}R_j^\alpha R_j^\beta =-\frac 2M%
Tr_{\omega L}\sum\limits_{{\bf k}}\left( \Sigma ^s\frac{\partial G^{\uparrow
}\left( {\bf k}\right) }{\partial k\alpha }\Sigma ^s\frac{\partial
G^{\downarrow }\left( {\bf k}\right) }{\partial k_\beta }\right)   \label{D2}
\end{equation}
where ${\bf k}$ is the quasimomentum and the summation is over the Brillouin
zone. In cubic crystals $D_{\alpha \beta }=D\delta _{\alpha \beta }$. For
arbitrary ${\bf q}$, the expression of magnon spectrum in terms of $J_{ij}$
is valid only in the adiabatic approximation, i.e. provided that the magnon
frequencies are small in comparison with characteristic electronic energies.
Otherwise, collective magnetic excitations which are magnons cannot be
separated from non-coherent particle-hole excitations (Stoner continuum) 
\cite{moriya} and magnon frequencies (\ref{om}) are not the exact poles of
transverse magnetic susceptibility (which are even not real at large ${\bf q}
$).

Now we have to consider the accuracy of expressions for $J_{ij}$ (\ref{Jij})
themselves. Eqs. (\ref{torque}) and (\ref{torque2}) are exact in LDA++
approach (i.e. with the only assumption about the {\it local} self-energy).
Hence, if one postulate the existence of effective spin Hamiltonian in the
sense of Eq.(\ref{torque3}), Eq. (\ref{Jij}) is also exact. However, they do
not have rigorous connection with the static transverse spin susceptibility.
The latter is expressed in terms of the matrix 
\[
\frac{\delta ^2\Omega _{}}{\delta \varphi _i^\alpha \delta \varphi _j^\beta }%
=\frac{\delta V_i^\alpha }{\delta \varphi _j^\beta }
\]
{\it without }the restriction $G=const.$ They differ from our exchange
parameters by the terms containing $\frac{\delta \widehat{G}}{\delta {\bf %
\varphi }_i}.$ From the diagrammatic point of view in the framework of DMFT 
\cite{dmft} they are nothing but the vertex corrections. We do not present
the corresponding expressions since the benefits of the introducing of
exchange parameters beyond adiabatic approximation which is equivalent to (%
\ref{torque3}) is doubtful. In more rigorous consideration it is convenient
to work directly with DMFT expressions for static and dynamic susceptibility 
\cite{dmft}. 

Thus one can see that, generally speaking,the exchange parameters differ
from the exact response characteristics defined via static susceptibility
since the latter contains vertex corrections. At the same time, our
derivation of exchange parameters seems to be rigorous in the adiabatic
approximation for spin dynamics when spin fluctuation frequency is much
smaller than characteristic electron energy. The situation is similar to the
case of electron-phonon interactions where according to the Migdal theorem
vertex corrections are small in adiabatic parameter (ratio of characteristic
phonon energy to electron one)\cite{migdal}. The derivation of the exchange
parameters from the variations of thermodynamic potential, being
approximate, can be useful nevertheless for the fast and accurate
calculations of different magnetic systems.

Note that in the LDA++ approach, as well as in LDA+U method and in contrast
with usual LSDF, one can rotate separately spins of states with given
orbital quantum numbers $L,L^{\prime }$ . For example, for non-relativistic
case one can obtain 
\[
\Omega _{spin}=-\sum_{iL,jL^{\prime }}J_{iL,jL^{\prime }}{\bf e}_{iL}{\bf e}%
_{jL^{\prime }} 
\]
where 
\[
J_{iL,jL^{\prime }}=-Tr_\omega \left( \Sigma _{iL}^sG_{iL,jL^{\prime
}}^{\uparrow }\Sigma _{jL^{\prime }}^sG_{jL^{\prime },iL}^{\downarrow
}\right) 
\]
are orbital dependent exchange parameters.

\subsection{Magnetic anisotropy}

Let us consider now the change of spin energy at the rotation of all the
spins at the same angle. It is definitely zero in nonrelativistic case. In
the presence of spin-orbit coupling, it is nothing but the energy of
magnetic anisotropy. One can obtain from Eq.(\ref{Hamilt}) 
\begin{eqnarray}
\Omega _{anis} &=&Tr_{\omega L}\sum\limits_{ij}\left\{ \left( {\bf G}%
_{ij}^s\times {\bf \Sigma }_j^s\right) \left( {\bf G}_{ji}^s\times {\bf %
\Sigma }_i^s\right) \right\} =  \label{anis} \\
&&\ \ Tr_{\omega L}\sum\limits_{{\bf k}}\left\{ \left( {\bf G}^s\left( {\bf k%
}\right) \times {\bf \Sigma }^s\right) \left( {\bf G}^s\left( {\bf -k}%
\right) \times {\bf \Sigma }^s\right) \right\}  \nonumber
\end{eqnarray}
where the last equality is valid for ferromagnets with one magnetic atom per
unit cell, {\bf k} is the quasimomentum and the summation is over Brillouin
zone.

Finally, note that we use essentially three properties of LDA++ approach:
(i) locality of self-energy (ii) spin-independence of bare Green function
(i.e. spin-independence of bare LDA spectrum; all magnetic effects including
Hartree-Fock ones are included in self- energy) and (iii) approximations for
self energy have to be conserving, or ``$\Phi $- derivable'' since only in
that case analog of ``local force theorem'' (\ref{var4}) takes place.

\section{Exchange interactions in ferromagnetic iron}

\subsection{Computational technique}

As an example we calculate the magnetic properties of ferromagnetic iron
using the most accurate method to take into account local correlations. For
this purpose we use the local quantum Monte-Carlo approach\cite{dmft} with
the generalization to the multiband case\cite{MQMC}.

We start from LDA+U Hamiltonian in the diagonal density approximation: 
\begin{eqnarray}
H &=&\sum_{\{im\sigma \}}t_{im,i^{\prime }m^{\prime }}^{LDA}c_{im\sigma
}^{+}c_{i^{\prime }m^{\prime }\sigma }+  \label{calc1} \\
&&\frac 12\sum_{imm^{\prime }\sigma }U_{mm^{\prime }}^in_{im\sigma
}n_{im^{\prime }-\sigma }+  \nonumber \\
&&\frac 12\sum_{im\neq m^{\prime }\sigma }(U_{mm^{\prime }}^i-J_{mm^{\prime
}}^i)n_{im\sigma }n_{im^{\prime }\sigma }  \nonumber
\end{eqnarray}
where $t^{LDA}$ is effective single-particle Hamiltonian obtained from the
non-magnetic LDA with the corrections for double counting of the average
interactions among d-electrons; $i$ is the site index and $m$ is the orbital
quantum numbers; $\sigma =\uparrow ,\downarrow $ is the spin projection; $%
c^{+},c$ are the Fermi creation and annihilation operators ($n=c^{+}c$).

The screened Coulomb and exchange vertex for the d-electrons{\ 
\begin{eqnarray}
U_{mm^{\prime }} &=&<mm^{\prime }|V_{scr}^{ee}({\bf r-r}^{\prime
})|mm^{\prime }>  \label{calc2} \\
J_{mm^{\prime }} &=&<mm^{\prime }|V_{scr}^{ee}({\bf r-r}^{\prime
})|m^{\prime }m>  \nonumber
\end{eqnarray}
are expressed via the effective Slater integrals and corresponds to the
average }$U=2.3$ eV and $J=0.9$ eV\cite{lda++}. We use the minimal $spd$%
-basis in the LMTO-TB formalism and numerical orthogonalization for $t^{LDA}(%
{\bf k})$  matrix \cite{lda++}. The Matsubara frequencies summation in our
calculations corresponds to the temperature of about T=850 K.

Local Green-function matrix has the following form 
\begin{equation}
G(i\omega )=\sum_{{\bf k}}\{i\omega +\mu -t^{LDA}({\bf k})-\Sigma (i\omega
)\}^{-1}  \label{Gk}
\end{equation}

Note that due to cubic crystal symmetry of ferromagnetic bcc-iron the local
Green function is diagonal both in the orbital and the spin indices and the
bath Green function is defined as 
\begin{equation}
G_m^0(i\omega )=G_m(i\omega )+\Sigma _m(i\omega )  \label{Sig}
\end{equation}

The local Green functions for the imaginary time interval $\left[ 0,\beta
\right] $ with the mesh $\tau _l=l\Delta \tau $, $l=0,...,L-1,$ and $\Delta
\tau =\beta /L$, where $\beta =\frac 1T$ is calculated in the path-integral
formalism\cite{dmft,MQMC}:{\ 
\begin{equation}
G_m^{ll^{\prime }}=\frac 1Z\sum_{s_{mm^{\prime }}^l}\det
[O(s)]*G_m^{ll^{\prime }}(s)  \label{QMC}
\end{equation}
here we redefined for simplicity $m\equiv \{m,\sigma \},Z$ is the partition
function and the so-called fermion-determinant }$\det [O(s)]$ and the Green
function for arbitrary set of the auxiliary fields{\ }$G(s)=O^{-1}(s)$ {are
obtained via the Dyson equation\cite{hirsch} for imaginary-time matrix} $(%
{\bf G}_m(s)\equiv G_m^{ll^{\prime }}(s))$: 
\[
{\bf G}_m=[{\bf 1}-({\bf G}_m^0-{\bf 1})(e^{V_m}-{\bf 1})]^{-1}{\bf G}_m^0 
\]
{where the effective fluctuation potential from the Ising fields }$%
s_{mm^{\prime }}^l=\pm 1$ is

\begin{figure}[tbp]
\vskip -.0cm \centerline{\ \epsfig{file=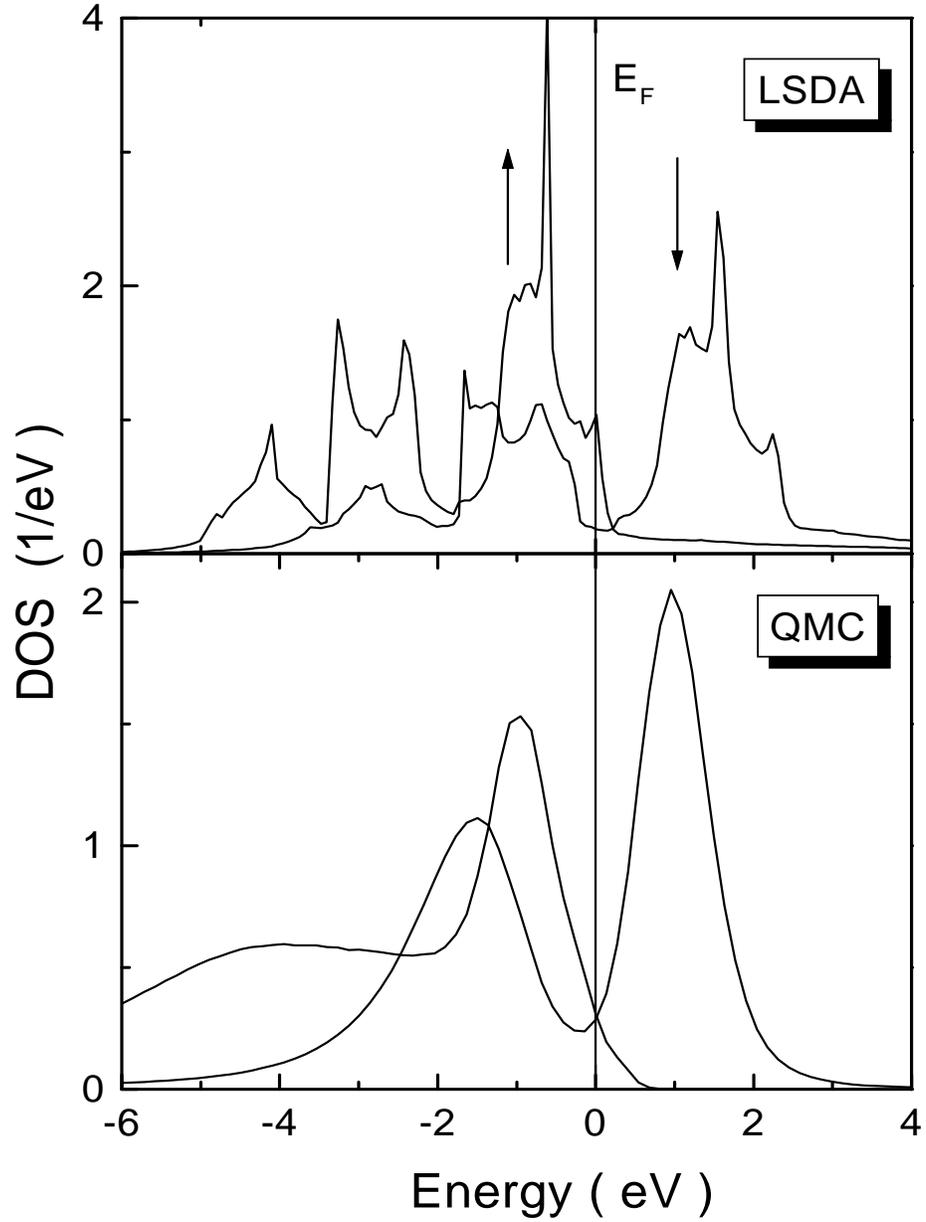, width=14cm,
height=19cm,angle=0} } \vskip 2.0cm
\caption{ Total spin-polarized density of states for ferromagnetic iron in
the LSDA and LDA+QMC approximations.}
\label{DOS}
\end{figure}

{\ 
\begin{eqnarray*}
V_m^l &=&\sum_{m^{\prime }(\neq m)}\lambda _{mm^{\prime }}s_{mm^{\prime
}}^l\sigma _{mm^{\prime }} \\
\sigma _{mm^{\prime }} &=&\{ 
\begin{array}{c}
1,m<m^{\prime } \\ 
-1,m>m^{\prime }
\end{array}
\end{eqnarray*}
and the discrete Hubbard-Stratonovich parameters are $\lambda _{mm^{\prime
}}={\rm arccosh}[\exp (\frac 12\Delta \tau U_{mm^{\prime }})] $ \cite{hirsch}%
. The main problem of the multiband QMC formalism is the large number of the
auxiliary fields }$s_{mm^{\prime }}^l.$ For each time slices $l $ it is
equals to $M(2M-1)$ where$M$ is the total number of the orbitals which is
equal to 45 for d-states. We compute the sum over this auxiliary fields in
Eq.\ref{QMC} using important sampling QMC algorithm and performed a dozen of
self-consistent iterations over the self-energy Eqs.(\ref{Gk},\ref{Sig},\ref
{QMC}). The number of QMC sweeps was of the order of 10$^5$ on the CRAY-T3e.
The final $G_m(\tau )$ has very little statistical noise. We use maximum
entropy method\cite{MEM} for analytical continuations of the QMC Green
functions to the real axis. Comparison of the total density of states with
the results of LSDA calculations (Fig.\ref{DOS}) shows a reasonable
agreement for single-particle properties of not ``highly correlated''
ferromagnetic iron.

Using the self-consistent values for $\Sigma _m(i\omega )$ we calculate the
exchange interactions (Eq.\ref{Jq}) and spin-wave spectrum (Eq.\ref{om})
using the four-dimensional fast Fourier transform (FFT) method\cite{FFT} for 
$({\bf k},i\omega )$ space with the mesh $20^3\times 320$. We compare the
results for the exchange interactions with corresponding calculations for
the LSDA method\cite{LKG}.

\subsection{Computational results}

\begin{figure}[tbp]
\vskip -.0cm \centerline{\ \epsfig{file=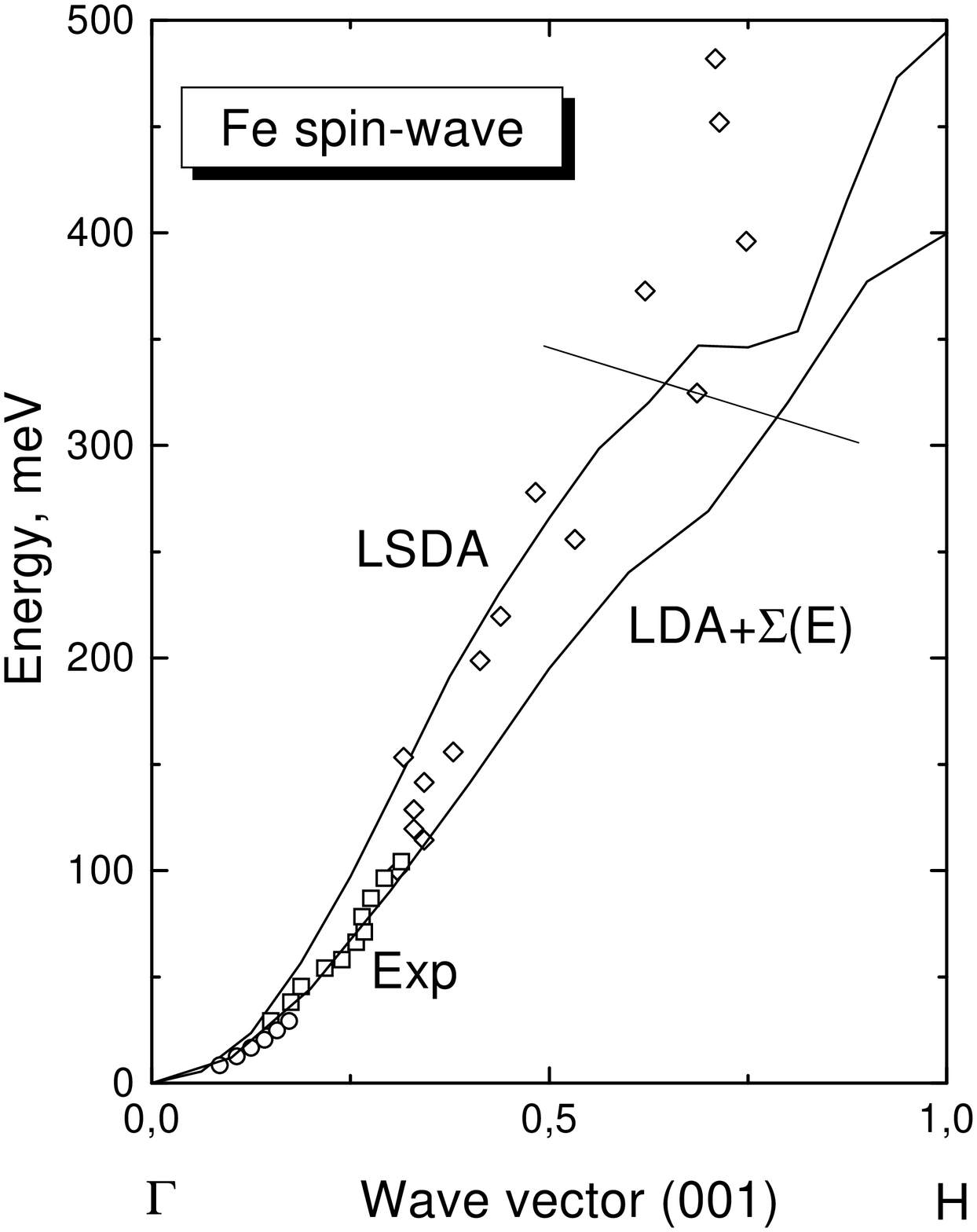, width=9cm,
height=12cm,angle=0} \epsfig{file=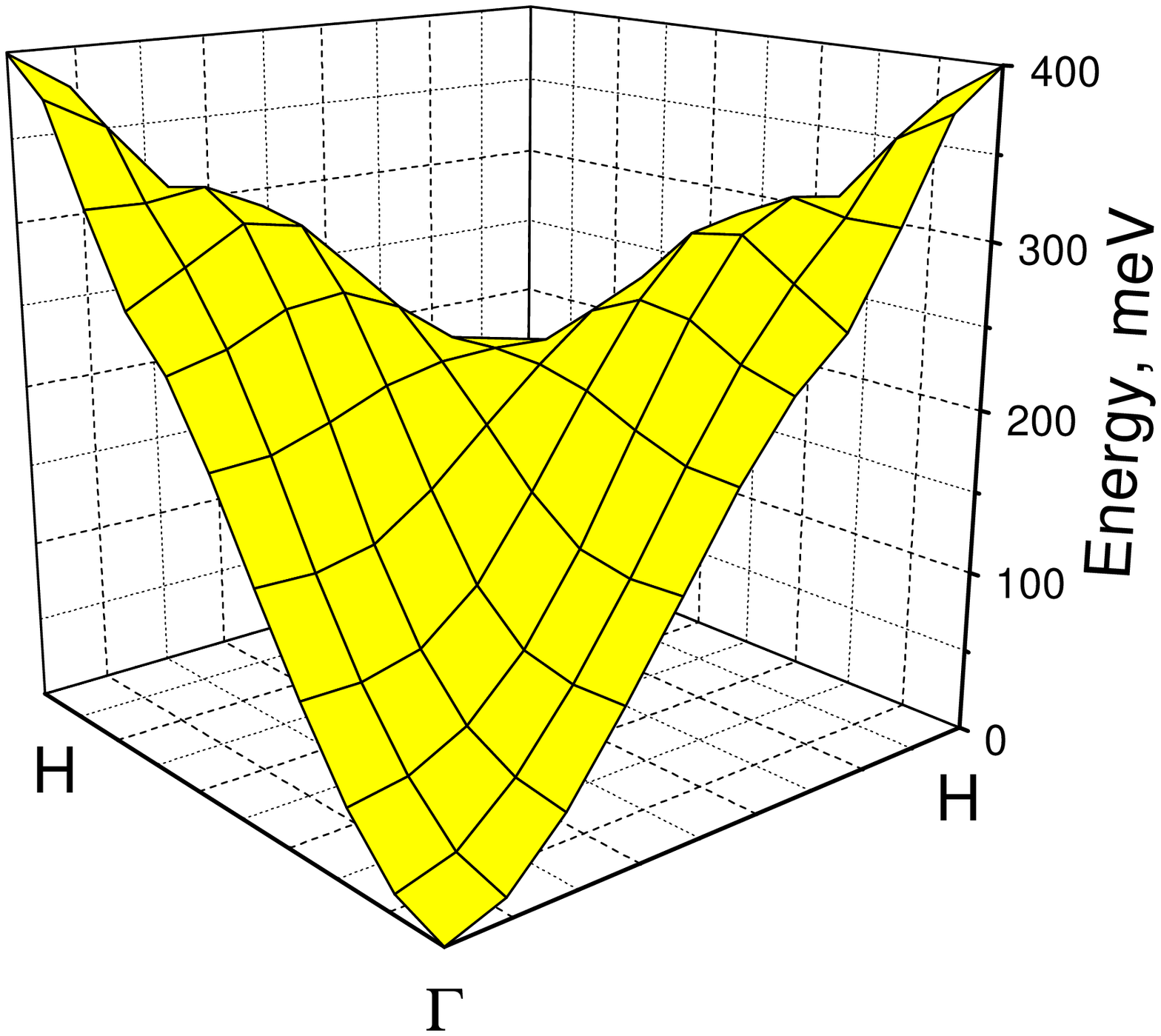, width=9cm, height=12cm,angle=0} }
\vskip 1.0cm
\caption{The spin-wave spectrum for ferromagnetic iron in the LSDA and LDA+$%
\Sigma$ approximations compared with different experiments (circles [16],
squares [17], and diamonds [18]) (a); The corresponding spin-wave spectrum
from LDA+$\Sigma$ scheme in the (110) plane (b).}
\label{SW}
\end{figure}

The spin-wave spectrum for ferromagnetic iron is presented in Fig.\ref{SW}
in comparison with the results of LSDA-exchange calculations and with
different experimental data\cite{omexp1,omexp2,omexp3} . Note that for
high-energy spin-waves the experimental data\cite{omexp3} has large
error-bars due to Stoner damping (we show one experimental point with the
uncertainties in the ``$q$'' space). On the other hand, the expression of
magnon frequency in terms of exchange parameters itself becomes problematic
in that region due to breakdown of adiabatic approximation, as it is
discussed above. Therefore we do not believe that good agreement of LSDA
data with the experimental ones for large wave vectors is too important
since it may be a result of ''mutual cancellation'' of inaccuracies of
adiabatic approximation and LSDA itself. At the same time, our lower-energy
spin-waves spectrum (where the theory is reliable) agree better with the
experiments then the result of LSDA calculations. 
Our LSDA spin-wave spectrum agree well with the results of
frozen magnon calculations \cite{Sandratskii,Halilov}.
 Note that in the LSDA
scheme one could  use the linear-response formalism\cite{Savr} to calculate
the spin-wave spectrum with the Stoner renormalizations, which should gives
in principle the same spin-wave stiffness as our LSDA calculations. The
corresponding exchange parameters and spin-waves stiffness (Eq.\ref{D})  are
presented in the Table I. The general trend in the distance dependence of
exchange interactions in ferromagnetic iron is similar in both schemes, but
relative strength of various interactions is quite different. Experimental
value of spin-wave stiffness D=280 meV/A$^2$\cite{omexp2} agrees well with
the theoretical LDA++ estimations.

\section{Conclusions}

In conclusion, we present a general method for the investigation of magnetic
interactions in the correlated electron systems. This scheme is not based on
the perturbation theory in ``$U$'' and could be applied for rare-earth
systems where both the effect of the band structure and the multiplet
effects are crucial for a rich magnetic phase diagram. Our general
expressions are valid in relativistic case and can be used for the
calculation of both exchange and Dzialoshinskii- Moriya interactions, and
magnetic anisotropy. An illustrative example of ferromagnetic iron shows
that the correlation effects in exchange interactions may be noticeable even
in such moderately correlated systems. For rare-earth metals and their
compounds, colossal magnetoresistance materials or high-$T_c$ systems, this
effect may be much more important. For example, the careful investigations
of exchange interactions in $MnO$ within the LSDA, LDA+U and optimized
potential methods for $MnO$ \cite{Sol} show the disagreement with
experimental spin-wave spectrum (even for small ${\bf q}$) , and indicate a
possible role of correlation effects.

As for the formalism itself, this work demonstrates an essential difference
between spin density functional approach and LDA++ method. The latter deals
with the thermodynamic potential as a functional of Green function rather
than electron density. Nevertheless, there is a deep formal correspondence
between two techniques (self-energy corresponds to the exchange- correlation
potential, etc). In particular, an analog of local force theorem can be
proved for LDA++ approach. It may be useful not only for the calculation of
magnetic interactions but also for elastic stresses, in particular,
pressure, and other physical properties.

\section{Acknowledgments}

The calculations were performed on Cray T3E
computers in the Forschungszentrum J\"ulich with grants of CPU time from the
Forschungszentrum and John von Neumann Institute for Computing (NIC).
This work was partially supported by Russian Basic Research Foundation,
grant 98-02-16279

\newpage

\begin{table}[tbp]
\caption{ Parameters of exchange interactions and spin-wave stiffness for
ferromagnetic iron calculated with the LSDA and LDA+$\Sigma $ methods. }
\label{JJ}\vskip 1.0cm 
\begin{tabular}{|l|lllllll|l|}
\hline
meV & J$_0$ & J$_1$ & J$_2$ & J$_3$ & J$_4$ & J$_5$ & J$_6$ & D (meV/A$^2$)
\\ \hline
LSDA & 166.1 & 16.48 & 8.07 & 0.25 & -1.03 & -0.31 & 0.26 & 320 \\ \hline
LDA+$\Sigma $ & 115.8 & 13.31 & 2.5 & 0.73 & -0.38 & -0.83 & 0.01 & 260 \\ 
\hline
\end{tabular}
\end{table}

\newpage
\vskip 1.0cm 

\end{document}